\documentclass{aa}  
\usepackage{graphicx}
\usepackage{txfonts}
\begin{document}
   \title{SuperWASP observations of long timescale photometric variations in cataclysmic variables}

   \subtitle{}

   \author{N.L. Thomas\inst{1}
          \and
      A.J. Norton\inst{1}
          \and
          D.Pollacco\inst{2}
          \and
          R.G. West\inst{3}
      \and
      P.J. Wheatley\inst{4}
      \and
          B. Enoch\inst{5}
          \and
      W.I. Clarkson\inst{6}}

   \offprints{}

   \institute{ Department of Physics and Astronomy, The Open University,
              Milton Keynes, MK7 6AA, UK\\
              \and
              Astrophysics Research Centre, School of Mathematics and Physics, Queen's University, University Road, Belfast BT7 1NN, UK\\
              \and
              Department of Physics and Astronomy, University of Leicester, Leicester LE1 7RH, UK\\
              \and
              Department of Physics, University of Warwick, Coventry CV4 7AL, UK\\
              \and
          SUPA, School of Physics and Astronomy, University of St Andrews, North Haugh, St Andrews, Fife KY16 9SS\\
              \and
            STScI, 3700 San Martin Drive, Baltimore, MD 21218, USA\\}

   \date{Received 07 06 2009 / Accepted  26  01  2010}

   \abstract  
   {}
   {We investigated whether the predictions and results of Stanishev et al (2002) concerning a possible relationship between eclipse depths in PX And and its retrograde disc precession phase, could be confirmed in long term observations made by SuperWASP. In addition, two further CVs (DQ Her and V795 Her) in the same SuperWASP data set were investigated to see whether evidence of superhump periods and disc precession periods were present and what other, if any, long term periods could be detected.}
   {Long term photometry of PX And, V795 Her and DQ Her was carried out and Lomb-Scargle periodogram analysis undertaken on the resulting light curves. For the two eclipsing CVs, PX And and DQ Her, we analysed the potential variations in the depth of the eclipse with cycle number.}
   {The results of our period and eclipse analysis on PX And confirm that the negative superhump period is 0.1417 $\pm$ 0.0001d.We find no evidence of positive superhumps in our data suggesting that PX And may have been in a low state during our observations. We improve on existing estimates of the disc precession period and find it to be 4.43 $\pm$ 0.05d. Our results confirm the predictions of Stanishev et al (2002). We find that DQ Her does not appear to show a similar variation for we find no evidence of negative superhumps or of a retrograde disc precession. We also find no evidence of positive superhumps or of a prograde disc precession and we attribute the lack of positive superhumps in DQ Her to be due to the high mass ratio of this CV. We do however find evidence for a modulation of the eclipse depth over a period of 100 days which may be linked with solar-type magnetic cycles which give rise to long term photometric variations. The periodogram analysis for V795 Her detected the likely positive superhump period 0.1165d, however, neither the 0.10826d orbital period nor the prograde 1.53d disc precession period were seen. Here though we have found a variation in the periodogram power function at the positive superhump period, over a period of at least 120 days.}
   {}
              
\maketitle
\section{Introduction}

Several kinds of cataclysmic variables (CVs) exhibit photometric modulations near to (but generally not precisely at) the binary's orbital period known as superhump variations. Positive superhumps (also called common or normal superhumps) are seen in the short-period CV systems known as the SU UMa dwarf novae during so-called superoutbursts (see e.g. Patterson \cite{Patterson1998}, Warner \cite {Warner}). The typical values of the peak-to-trough amplitude of the superhumps are about 20-40 percent of the superoutburst amplitude. Normal superhumps oscillations can be characterised by the positive superhump excess period $\varepsilon_+$ as
 \begin{math}\varepsilon_+=(\emph{P}_{sh+} - \emph{P}_{orb})/\emph{P}_{orb}\end{math} 
  where $\emph{P}_{sh+}$ and $\emph{P}_{orb}$ are the superhump and orbital periods respectively.
$\varepsilon_+$ is found to be a function of mass ratio (Baker-Branstetter \& Wood \cite{Baker}, Patterson \cite{Patterson2001}, Patterson et al \cite{Patterson2002}, \cite{Patterson2005}, and Montgomery \cite{Montgomery}). There is also a trend of increasing positive excess of the superhump period over the orbital one with the orbital period (Stolz \& Schoembs \cite{Stolz1984}, Patterson \cite{Patterson1999}).

The currently accepted model explaining common superhumps is that they result from tidal stressing of the accretion disc by the secondary star in which the disc encounters the 3:1 ecentric inner Linblad resonance with the tidal potential of the secondary star, thereby becomes eccentric, and precesses (see Osaki \cite{Osaki1985}; Whitehurst \cite{Whitehurst1988a},  \cite{Whitehurst1988b}; Hirose \& Osaki \cite{Hirose1990}; Lubow \cite{Lubow1991a}, \cite{Lubow1991b}; Whitehurst \& King \cite{Whitehurst1991}; Whitehurst \cite{Whitehurst1994}; Murray\cite{Murray}). It would appear e.g. Smith et al (\cite{Smithetal2007}) that the region in the disc overwhelmingly responsible for generating dissipation-powered superhumps is where the 3:1 resonance is the most dominant, while the 4:1 and 5:1 ecentric inner Linblad resonances (Lubow \cite{Lubow1991b}) may play a role in the dynamics of eccentric discs for mass ratios $q < 0.24$. Theoretically it is  found (Paczynski \cite{Paczynski1977}) that there is an upper limit on the mass ratio of $q < 0.25$ which will allow apsidal precession to occur and enable positive superhumps to form.Observationally (Patterson et al \cite{Patterson2005}) the limit is seen to be nearer $q = 0.3$. \\ The general model of superhump formation may be seen in the hydrodynamic particle simulations of Hirose and Osaki (\cite{Hirose1990}), Simpson \& Wood (\cite{Simpson1998}), Wood et al (\cite{Woodetal2000}) and by Smith et al (\cite{Smithetal2007}) etc who show it is the luminosity arising from energy dissipation in enhanced spiral density waves which extend radially through the disc that gives rise to the positive superhump lightcurves. Because the disc is precessing in a prograde direction, the secondary meets up with the line of apsides on a period slightly longer than the orbital period. This longer period is the apsidal superhump period $P_{\rm sh+}$ and the two periods are related to the prograde precession period by
\begin{math}1/P_{\rm prec+}= 1/P_{\rm orb} - 1/P_{\rm sh+}\end{math}.

While positive superhumps have periods of a few percent longer than the orbital period, some CVs, however, show negative superhumps (sometimes called infrahumps) i.e. modulations at periods that are a few per cent shorter than the binary orbital period. They are more rare than positive superhumps with {$\sim31$} systems currently known (Montgomery (\cite{Montgomery2009b}). In a few systems they appear simultaneously with positive superhumps (eg Retter et al \cite{Retter2001}), but in other systems they are the only kind observed, and alternations between the positive and negative superhumps have been observed in a few objects (eg Skillman et al \cite{Skillmanetal1998}).
Retter et al (\cite{Retter2001}) has shown that for those systems that display both negative and positive superhumps the relationship between the negative superhump deficit $\varepsilon_-$ and the positive superhump excess $\varepsilon_+$ shows a clear trend with orbital period. In contrast to positive superhumps, the occurence of negative superhumps is not confined by a particular mass ratio for they are found in both short period systems such as AM CVn and long period ones like TV Col.

Although there are competing theories for the cause of negative superhumps (see Montgomery \cite{Montgomery2009a} \& \cite{Montgomery2009b} for a review and references therein), they are generally thought to arise from the slow retrograde precession of an accretion disc which is tilted out of the orbital plane. Even here there is no consensus on the the exact mechanism which powers the negative superhump light source. For example Wood et al (\cite{Woodetal2009}) suggest, from their SPH simulations, that it is the transit of the accretion stream impact spot across the face of the disc that may be the source of the negative superhumps. However, interestingly, their simulations showed that removal of of the gas stream (and hence the bright spot) does not get rid of the negative superhumps. Montgomery (\cite{Montgomery2009a}) on the other hand suggests, again from SPH modeling, that the source of the negative superhumps is additional light from the innermost disc annuli, and this additional light waxes and wanes with the amount of gas stream overflow received as the secondary orbits. Montgomery shows that a minimum disc tilt angle of 4\degr is required to produce negative superhumps and that as the angle of tilt increases the higher their amplitude. According to Smak (\cite{Smak2009}) the source to the disc tilt lies in stream-disc interactions arising from differential irradiation of the secondary star. The origin of the retrograde precession remains under debate with competing theories ranging from the freely precessing disc model of Bonnet Bidaud et al (\cite{Bonnet1985}) to gravitational effects similar to those on the Earth Moon system which cause the Earth to precess retrogradley (Patterson et al \cite{Pattersonetal1993}, Montgomery \cite{Montgomery2009b}).

In those systems which display negative superhumps, the tilted disc is precessing in a retrograde direction and the secondary meets up with the line of nodes on a period slightly shorter than the orbital period. This shorter period is the nodal superhump period $\emph{P}_{sh-}$ and the two periods are related to the the retrograde precession period by
\begin{math}1/P_{\rm prec-}= 1/P_{\rm sh-} - 1/P_{\rm orb}\end{math}. Similarly the negative superhump deficit,  $\varepsilon_-$, is defined in an analogous way to the relationship for the superhump excess as,
\begin{math}\varepsilon_-=(P_{\rm sh-} - P_{\rm orb}) / P_{\rm orb}\end{math}.

Tilted discs have observational consequences other than the superhump period itself. These can be used to establish that a tilted accretion disc may be present (with or without spectroscopic signatures), and to constrain the properties of such a configuration. Stanishev et al (\cite{Stanishev}) undertook a photometric study of the SW Sex novalike PX And and they performed a periodogram analysis of the observations obtained in October 2000. This revealed the presence of three signals, a 0.142d period which corresponds to negative superhumps, a period of 4.8d for the retrograde precession period of the accretion disc, and a 0.207d period, the origin of which was unknown. Stanishev et al (\cite{Stanishev}) found that the mean out of eclipse magnitude showed large variations, modulated with a period of about 5 days and a full amplitude of approximately 0.5 mag, which they modelled with a sinusoid. Their observations therefore suggested that the eclipse depths were modulated with the disc precession cycle. In order to verify this they subtracted the best-fit sinusoid from the mean out of eclipse flux, which showed that the eclipses are deepest at the minimum of the precession cycle. The mean eclipse depth was found to be approx 0.89 mag and the amplitude of the variation to be approx 0.5 mag. Based on the observed amplitude of the 4.8d modulation, Stanishev at al (\cite{Stanishev}) estimated PX And's disc tilt angle to be between 2.5\degr and 3\degr depending upon the assumed system inclination.

\section{Aims of this work} 
The object of this study was to investigate whether the predictions and results of Stanishev et al (\cite{Stanishev}) concerning a possible relationship between the depth of the eclipses in PX And and its negative superhump period could be confirmed in longer term observations obtained by SuperWASP. In addition, two further CVs (DQ Her and V795 Her) in the same SuperWASP data set were investigated. For the eclipsing CV DQ Her the object of the analysis was to look for a possible retrograde precession period that is associated with negative superhumps, search for a prograde precessional period that is indicated by the presence of positive superhumps and determine what, if any, other long term periods could be seen. As V795 Her is reported to have an orbital period of 0.10826d and a superhump period  0.1165d (Casares et al \cite{Casares1996}) the aim was to establish whether there is any sign of the prograde disc precession period at 1.53d in the SuperWASP observations.

\section{The target stars}
\subsection{PX And}
PX Andromedae is a nova-like CV having an orbital period of 0.14635d (3.51 h) that displays permanent negative superhumps, having a reported superhump period of $\backsim 0.142d $ (3.41h) and a retrograde disc precession period of 4.4366d (Stanishev et al \cite{Stanishev}). In addition (Patterson \cite{Patterson1999}) found PX And to display a positive superhump period of 0.1595(2)d (3.83h). PX And is currently thought to be a member of the sub-class of CVs known as SW Sextantis stars (Hellier \cite{Hellier2000}) showing the most complicated behaviour of that group. The history of its period measurements is as follows. 
Thorstensen et al (\cite{Thorstensenetal1991}) in undertaking a spectroscopic study of PX And
calculated that the inclination of the system is approximately $74^{\circ}$ and found that it showed shallow eclipses which were highly variable in their eclipse depth of around 0.5 mag.  They determined an orbital period of 0.146353d, on the basis of their determination of Hα emission-line radial velocities; Hellier \& Robinson (\cite{HellierRobinson1994}) later undertook photometry of PX And and refined its ephemeris. Patterson (1999) reported PX And to show the presence of positive superhumps, with a superhump excess $\varepsilon_+$ of 0.089, and a period of 0.1595(2)d as well as a retrograde precession cycle with a period of 4 to 5 days.
Stanishev et al (2002) further investigated the system, as described in section 1, and improved the orbital ephemeris for PX And as
\begin{math}\emph{T}_{\rm min} [HJD] = 49238.833662(14) + 0.146352739(11) E \end{math}

Stanishev et al (\cite{Stanishev}) concluded that the relatively shallow eclipses show that the accretion disc in PX And is not totally eclipsed. Subsequently, Boffin et al (\cite{Boffin2003}) went on to further discuss the potential cause of the eclipse depth variations and introduced the possibility that the accretion disc in PX And may be warped.

\subsection{DQ Her}
The eclipsing binary system DQ Her consists of a cool star with a mass near 0.4~M$_{\odot}$ and a white dwarf with mass near 0.6~M$_{\odot}$ (Horne et al \cite{Horneetal1993}). 
DQ Her is now known to be an intermediate polar and is the prototype for a CV in which the white dwarf is rapidly rotating and has a strong magnetic field. 
Long-term variations in the brightness of DQ Her have been reported. Patterson et al (\cite{Pattersonetal1978}) presented evidence for a long-term period in DQ Her based on the O-C diagram for eclipse times indicating a possible 13 year periodicity, based on 1.5 cycles of data. A similar periodicity in the star's brightness was seen (Dmitrienko \cite{Dmitrienko}), which appeared to be correlated with orbital changes. The mean change of magnitude was reported as 0.6 mag. However unpublished data by Richman et al (\cite{Richmanetal1994}) based on a few dozen measurements suggest that any such modulation in brightness is smaller than the 0.6 mag claimed effect. Dai \& Qian (\cite{DaiQian}) derived a $13 \pm 3$ year variation, but
argued that this may be attributable to a third body in orbit around DQ Her such as a brown dwarf. They concluded that a detailed analysis of DQ Her's variation in brightness is necessary in order to confirm whether the  magnetic activity cycle on the red dwarf star in DQ Her is  strong enough to cause the observed modulation in the O-C diagram.
    DQ Her has an ephemeris  (Zhang et al \cite{Zhangetal1995}) of \\
 \begin{math}T_{\rm min} [HJD] = 2434954.94429 (\pm 8) + 0.1936208964(\pm 12) E \end{math}

\subsection{V795 Her}
    V795 Her is a Nova-like variable, possibly magnetic, and thought to be a SW Sex type star. It  is now known (Shafter et al \cite{Shafteretal1990}) to have an orbital period of 0.10826d (2.598h), and displays (Patterson et al \cite{Patterson1994}) permanent positive superhumps with a period of 0.11695d (2.80h). Note, however, the path to recognising this has been rather convoluted as described below.

V795 Her is one of the few cataclysmic variables with an orbital period lying near the middle of the gap in the distribution of periods between roughly 2 and 3 hrs (Whyte and Eggleton \cite{WhyteEggleton1980}; Robinson \cite{Robinson1983}). The system was first identified by Green et al (\cite{Greenetal1982}) who classified V795 Her (PG1711 +336) as a possible cataclysmic variable based on its emission line spectrum having high excitation lines and a UV excess.
Mironov et al (\cite{Mironovetal1983}) analysed 250 archival plates spanning 75 years, which showed that V795 Her exhibits slow brightness variations between B=13.2 and B=12.5 mag with no evidence for eruptions. They found a periodic variation in the light curve with full amplitude of about 0.2 mag and a period of 0.115883d (2.78h) with a possible alias period at 0.13117d  (3.148h). The light curve was found to have a saw-tooth shape. 
Observations by Baidak et al (\cite{Baidaketal1985}), who had more data than Mironov et al(\cite{Mironovetal1983}), subsequently derived a slightly different photometric period of 0.114488d (2.747h).
The 2.7h photometric period was initially thought to be the rotation period of the white dwarf. Thorstensen (\cite{Thorstensen1986}), who identified V795 Her to be a nova-like variable in a perpetual high state, obtained time resolved spectroscopic observations. However, since the emission lines in his spectra were relatively weak, his periodogram of the H$\alpha$ velocities was unable to confirm Baidek's photometric period of 0.114488d.  Interestingly, Thorstensen found the strongest frequency in his data was 1/0.63d i.e. a period of 1.6d close to the expected disc precession period of 1.53d. This was, however, rejected as there was evidence of line profile changes over some nights, and Thorstensen believing that that a higher frequency of 1/1.62d (= 0.62d period) was more credible, concluded that this 14.8h period may be the orbital period. Rosen et al (\cite{Rosenetal1989}) found two alternative values of the photometric period, 0.1157550d and 0.1158807d in their photometric and spectroscopic study of V795 Her. However, since their data sample was only 3.3 hrs in length, the precision of their spectroscopic period determination was insufficient to allow quantitative comparison with their photometric period.
    Shafter et al (\cite{Shafteretal1990}) undertook photometric and spectroscopic observation of V795 Her and established a spectroscopic period of 0.1164865(4)d from radial velocities measured in the wing of H$\alpha$, H$\beta$, and {He}II  $\lambda4686$ and identified this as the orbital period of the system. The dominant period in their data was, however, at approximately 0.108d, although one cycle per day aliases at 0.097d and 0.121d could not be excluded.
Ashoka et al (\cite{Ashokaetal1989}) determined the photometric period to be 0.115d, while Kalzuny (\cite{Kalzuny1989}) and Zhang et al (\cite{Zhangetal1991}) found a 0.11649d period. The 0.116d period was confirmed by Zwitter et al (\cite{Zwitteretal1994}) from their 3-day observations.
Patterson \& Skillman (\cite{PattersonSkillman1994}) found the 0.1158d periodic signal essentially disappeared during the period 1990-1994 and instead detected the 0.1086d orbital period from their photometric data.
Haswell et al (\cite{Haswelletal1994}) had a more extensive data set than Patterson \& Skillman but failed to detect a 1.5d - 1.6d period which would be expected for the prograde disc precession period, however this is probably because the photometric superhumps were known to be absent during the 1993 observations.
Casares et al (\cite{Casares1996}) undertaking optical photometry and spectroscopy confirmed the spectroscopic modulation of 0.1082d, which they attributed to the orbital period of the system. Their R band photometry was dominated by flickering and they found no evidence for modulation with the orbital period or a 0.1165d period. 

\section{Observations}
 The SuperWASP project (Pollacco et al \cite{Pollaccoetal2006}) is a wide field photometric survey designed to search for transiting exoplanets and other signatures of stellar variability on timescales from minutes to months. The SuperWASP telescopes (one on La Palma the other at the South African Astronomical Observatory) consist of an array of 8 CCD cameras fitted with Canon 200mm, f/1.8 telephoto lenses having an aperture of 11cm, fitted on a single fork mount. Observations of the three CVs were obtained from one of the 5 SuperWASP cameras available in 2004, each having a $7.8^{\circ}$ x $7.8^{\circ}$ field of view. The observations were unfiltered (white light) with the spectral transmission effectively defined by the optics, detectors and atmosphere. 

Long term photometry of PX And, V795 Her and DQ Her was undertaken during the first survey by the SuperWASP facility at the Observatorio del Roque de los Muchachos on La Palma in 2004. Norton et al (\cite{Nortonetal2007}) have previously presented a catalogue of all the SuperWASP objects from the 2004 observing season which displayed a periodic photometric modulation, and which were coincident with X-ray sources. The only three cataclysmic variables present in that sample were the three objects which form the basis of this study. All observations were taken between May and September 2004, the majority of these on a daily basis with the interval between successive observations being, on average, 6 minutes. However, there are several gaps in the data of length 2, 3, and 4 days where no observations were made. Photometry for each object was extracted using a 2.5 pixel aperture ($34^{\prime \prime}$~radius), and light curves obtained spanning 126 days containing 7467 data points, 138 days containing 4355 data points, and 150 days with 5000 data points for PX And, DQ Her, and V795 Her respectively.

\section{Data reduction}
The SuperWASP data reduction pipeline employs the same general techniques described by Kane et al (\cite{Kaneetal2004}) for the prototype WASP0 project. Bias frames, thermal dark-current exposures and twilight-sky flat-field exposures were secured and master bias, dark and flat-field frames were constructed and aperture photometry carried out. Data reduction was carried out, as described in detail by Pollacco et al (\cite{Pollaccoetal2006}), from raw instrumental to calibrated standard magnitudes. The effects of primary and secondary extinction, the instrumental colour response and the system zero-point were calculated to give calibrated magnitudes.
Approximately 100 bright, non-variable stars were adopted as secondary standards within each field. Their standardized magnitudes as determined over a few photometric nights were subsequently used to define the WASP V magnitude system for the field concerned. The mean SuperWASP magnitudes are defined as $2.5 \log_{10} (F/10^6)$ where $F$ is the mean SuperWASP flux in microVegas; it is a pseudo V magnitude that is comparable to the Tycho V magnitude.

\section{Periodogram and eclipse analysis}
 
Period determination for the three objects studied was undertaken by calculating Fourier transform power spectra of their SuperWASP light curves based on a modified Lomb-Scargle method (see Lomb \cite{Lomb1976} and Scargle \cite{Scargle1982}). The algorithm employed was optimised using the Horne and Baliunas (\cite{HorneBaliunas1986}) method which scales the periodogram by the total variance of the data. Period significance was determined by a Monte Carlo Permutation Procedure to calculate two complementary False Alarm Probabilities, and period errors were determined by calculating a 1$\sigma$ confidence interval on the dominant period using the Schwarzenberg-Czerny method (Schwarzenberg-Czerny \cite{Schwarzenberg-Czerny1991}).
   In order to determine whether there are aliases or false peaks in the periodic signals that are artefacts of the intervals between the observations, window functions were produced (Scargle \cite{Scargle1982}) which display the frequencies or periods between which the observations are made, and the relative number of observations at those particular frequency or period. Comparison of the window function with the periodogram enables the effects of the sampling rate to be identified and their impact on the results of the periodogram to be assessed.
We did not attempt to identify the nature of each prominent period detected, rather we looked to establish whether known or predicted periods could be confirmed in the periodograms. 
The raw data was initially visually scanned to identify where potential aliasing periods may arise. These were typically associated with the sampling rate of the raw observations which varied for each target or associated with the number and spread of eclipse determinations which could be made. Period ranges for the searches were selected in those regions which were thought either to contain known periodicities or potentially harbour periodicities related to the orbital period, superhump period and disc precession period. 

For the two eclipsing CVs PX And and DQ Her a period analysis of the variation in the depth of the eclipse with orbital cycle number or observation time was made. The depths of the eclipses were measured by the difference between the flux at the lowest point of the eclipse and the flux at a suitable reference position on the out of eclipse region. Since the out of eclipse regions were found to be highly variable from one orbital cycle to another, principally due to scatter in the data, different phase positions were examined and ultimately phase 0.5 was chosen as the reference since this phase consistently showed the least level of variation. Several methods of measuring the phase 0.5 flux were investigated including using 3 and 4 point moving averages and calculation of the median flux level between phase 0.45 and 0.55.   The 3 point moving average was found to be suitable for DQ Her where there was the least scatter in the data while the median flux level was found to be more suitable for PX And where there were relatively more data points with a large degree of scatter (see Fig.1). While the two techniques did give slightly different average values, this was not considered significant as to affect the results of the periodograms. 

\section{Results}
\subsection{PX And} 
 
\begin{figure}
   \resizebox{\hsize}{!}{\includegraphics{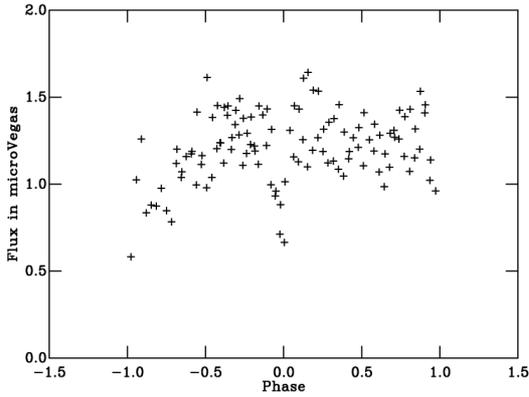}}
  \caption{Typical nightly lightcurve of PX And from SuperWASP data based on the orbital period taken from the Stanishev et al ephemeris.}
  \label{Fig.1}
\end{figure}

A Lomb-Scargle analysis was undertaken on the complete WASP data set for PX And. The initial period range chosen was 0.1d to 2.0d in order to see  what significant periods may be present in the data and if they show the presence of the orbital period or superhump periods. The  periodogram was set to 1500 steps in order to minimise the step size between periods (and hence maximising the number of potential periods which may be seen) while maintaining a reasonable computation process time. The orbital period of 0.1463d was clearly detected and  two possible candidate periods which could be the negative superhump period.  In addition the signature of a 0.207d period was seen as reported by Stanishev et al (\cite{Stanishev}). The corresponding window function shows peaks at 1d, 1/2d, 1/4d, 1/8d etc. These are likely to be associated with the sampling rate. 

\subsection{PX And superhump period}
In order to investigate the possible negative superhump period more closely we carried out a periodogram analysis  in the period range 0.1d to 0.3d. The result shows that the principal peak is at 0.1417d. This is likely to be the negative superhump period. However the window function shows a peak at 0.1429d ($\backsim1/7d$), which probably explains a neighbouring 0.1422d peak near the negative superhump period.
Thus it is likely that the superhump period is detected but sampling aliases confound the determination. In order to minimise the effects of the sampling rate, a contiguous subset of the PX And observations, 1137 observations over eight days between HJD 2453235 and 2453244 were analysed as follows:

Fig.2 shows the Lomb-Scargle analysis in the range 0.1 to 0.5 d  with 1500 trial periods for the PX And SuperWASP subset data. The maximum peak is the negative superhump period of 0.1417d $\pm$ 0.0001d. Other nearby peaks at 0.1243d $\pm$ 0.0005d, 0.1659d $\pm$ 0.0013 and 0.1986d $\pm$ 0.0027d are likely to be aliasing effects. The orbital period of 0.1463d was not seen in this small sample of data. We also did not find any evidence of the 0.1595d $\emph{P}_{sh+}$ period.

In order to confirm that 0.1417d is the principal period and the nearby peaks are simply aliases, the periodogram in Fig.2 was pre-whitened by removing the 0.1417d period and a period analysis carried out on the residuals. The result, shown in Fig.3, shows that the principal peaks around the negative superhump period have disappeared indicating that they were aliases of the negative superhump period.

The PX And SuperWASP data subset singly folded on the negative superhump period of 0.1417d (without binning) is shown in Fig.4. Despite the scatter in the data, the shape of the superhumps can be seen.

\begin{figure}
   \resizebox{\hsize}{!}{\includegraphics{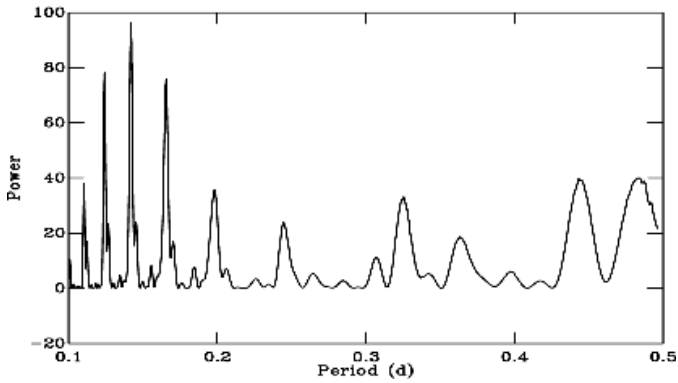}}
  \caption{PX And Subset data Lomb-Scargle periodogram with 1500 trial periods spanning 0.1 to 0.5 d. The highest peak in the figure is the negative superhump period at 0.1417d.}
  \label{Fig.2} 
\end{figure}
\begin{figure}
   \resizebox{\hsize}{!}{\includegraphics{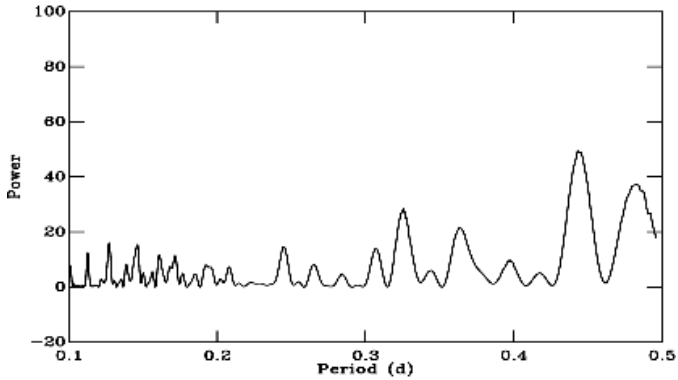}}
  \caption{PX And data subset for the region 0.1d to 0.5d with 1500 trial periods pre-whitened with 0.1417d period.}
  \label{Fig.3}
\end{figure}

\begin{figure}
   \resizebox{\hsize}{!}{\includegraphics{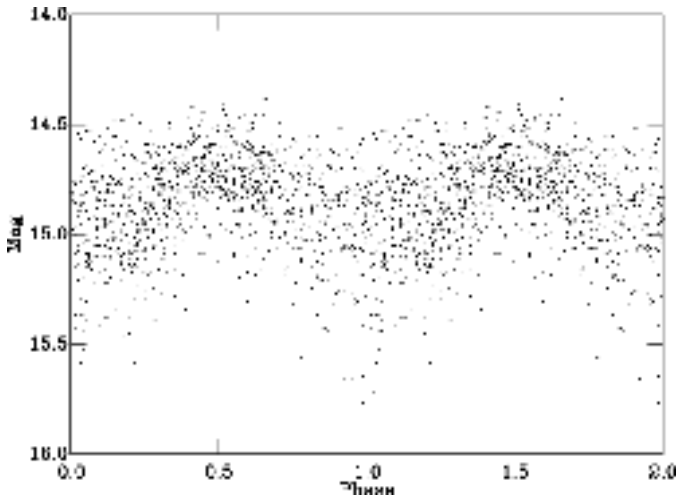}}
  \caption{PX And SuperWASP data subset folded on the negative superhump period of 0.1417d. The phase is repeated for clarity.}
  \label{Fig.4}
\end{figure}
\begin{figure}
   \resizebox{\hsize}{!}{\includegraphics{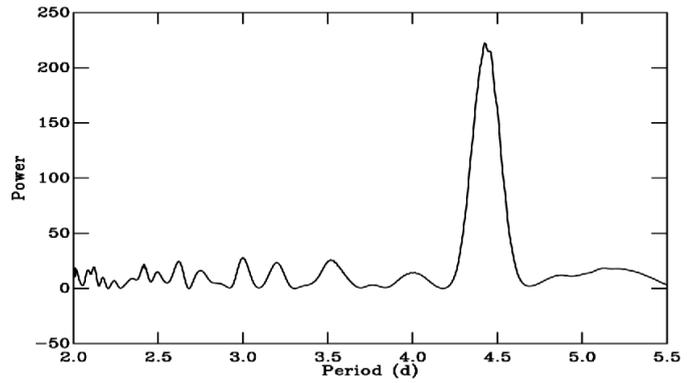}}
  \caption{PX And SuperWASP data periodogram spanning  2.0d to 5.5d with 1500 trial periods. The highest peak in the figure is the 4.43d retrograde disc precession period.} 
  \label{Fig.5}
\end{figure}

\begin{figure}
   \resizebox{\hsize}{!}{\includegraphics{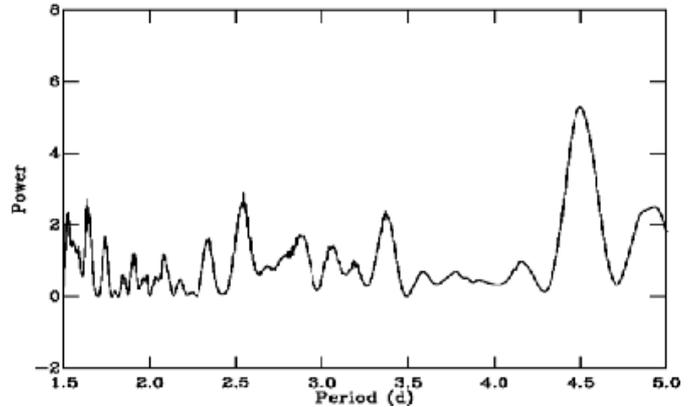}}
  \caption{PX And eclipse depth data periodogram spanning 1.5d to 5.0d periods with 1500 trial periods showing the retrograde disc precession period.}
  \label{Fig.6}
\end{figure}
\begin{figure}
   \resizebox{\hsize}{!}{\includegraphics{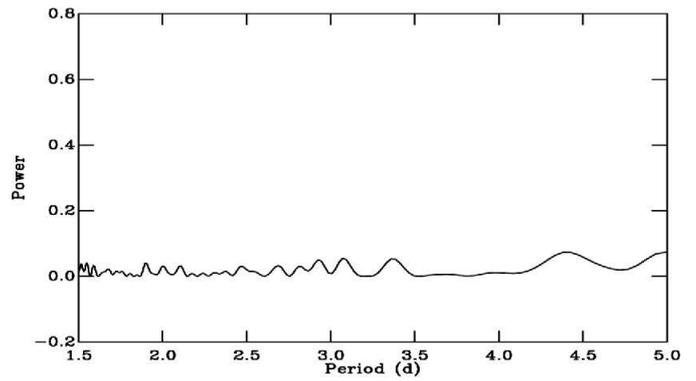}}
  \caption{PX And eclipse depth data spectral window for region 1.5d to 5.0d cycles.}
  \label{Fig.7}
\end{figure}
\begin{figure}
   \resizebox{\hsize}{!}{\includegraphics[scale=0.7]{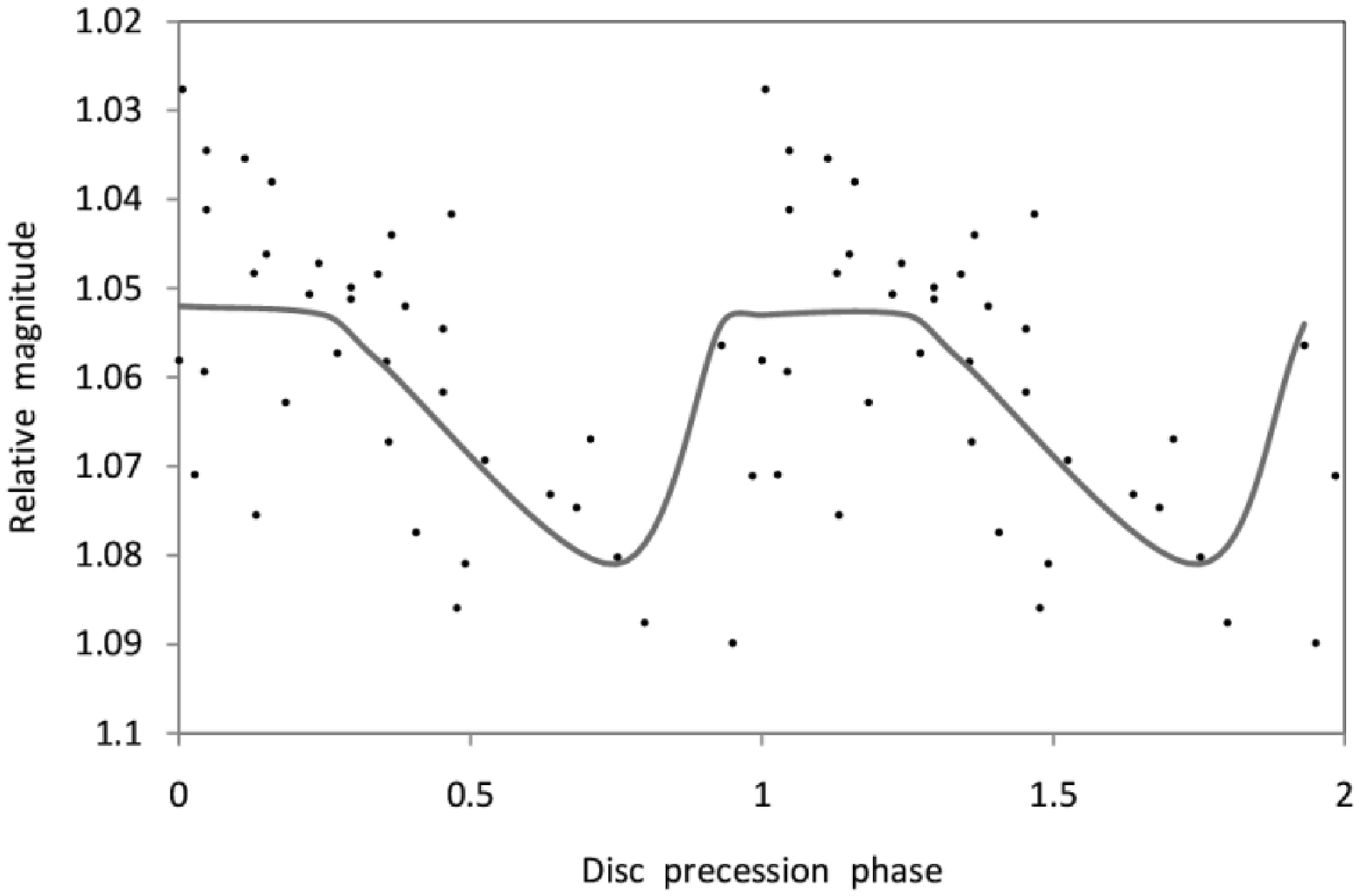}}
  \caption{PX And eclipse data folded on the negative precession period with a mean curve to the phase data using a spline interpolation. The peak around phase 1.0 is ignored as the curve is overplotted in order to aid the eye. The phase is repeated for clarity. }
  \label{Fig.8}
\end{figure}

\subsection{PX And disc precession period}  
A Lomb-Scargle Fourier analysis was performed on the complete PX And WASP data set in the period range 2d to 40.0d in 500 steps. The most prominent period was found to be 4.426d. Reducing the period range to 2.0d  to 5.5d and increasing the number of trial periods to 1500  improves the accuracy of the determination to give the prominent period as 4.43 $\pm$ 0.05d (Fig.5). 

The depth of the eclipses in PX And were determined by the method referred to in Section 4.5. Not all the eclipse data could be used because they were not amenable to satisfactory determination of their relative depth. Plotting the results against orbital cycle number  appears to show no obvious correlation between the eclipse depth and orbital cycle number. 

The out of eclipse flux  at  phase 0.5, which was used as the reference phase flux in the measurement  of eclipse depth, showed the least variation in flux in all the orbital phases measured, however it was not constant. In order to minimise the effect of the variation in the reference phase flux, the eclipse depth flux was measured as a fraction of the reference flux  The variation in eclipse depth  is of the order of $\pm$ 0.05 mag.

The relative eclipse flux data  was converted to relative magnitude and a Fourier analysis was performed in the region of 1.5d to 5.0d in order to see if the retrograde disc precession period was present. The results in Fig.6 show a definite pronounced peak in the region of 4.5d. The region around 4.0d shows no significant sampling effects around the retrograde disc precession period (see Fig.7). The Fourier analysis was repeated  for a shorter period range around this period confirming the peak at  4.43 $\pm$ 0.05d day is the retrograde disc precession period.

Finally Fig.8 shows the PX And eclipse depth data singly folded on the retrograde disc precession period with a mean curve to the phase data using a spline interpolation method. While there is a degree of scatter, the eclipse depth data clearly varies with the retrograde precession period, as suggested by Stanishev et al (\cite{Stanishev}).

\subsection{DQ Her}
 
While large photometric data sets enable a high degree of precision in determining periods, they also have the potential effect of introducing a large range of sampling aliases. Unlike the PX And data, the data for the other two CVs had a larger range of observation gaps. Therefore, in order to reduce the number of sampling aliases, a subset of the DQ Her SuperWASP data, consisting of 1770 observations  in a contiguous 35 day period between May 2004 and June 2004, was analysed to confirm the orbital period.

The results of the Lomb-Scargle analysis for the region between 0.1d and 2.0d are shown in Fig.9 together with the prominent periods in Table 1. The dominant period is 0.1936d $\pm$ 0.0001d, the orbital period. The window function for this period range in the data subset shows that the sampling rate in this data set is unlikely to affect the determination of the orbital period. 
\begin{figure}
   \resizebox{\hsize}{!}{\includegraphics{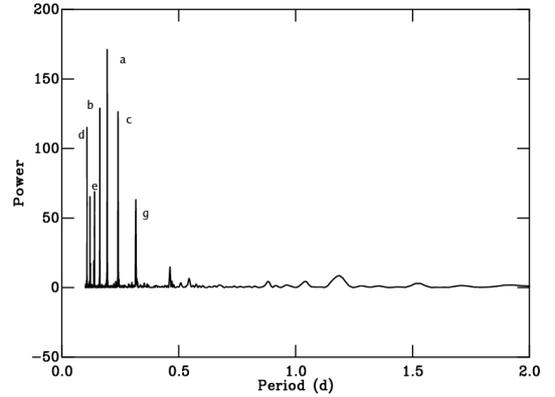}}
  \caption{DQ Her data subset periodogram for 0.1d to 2.0d  with 2000 trial periods. The dominant peak is the orbital period.}
  \label{Fig.9}
\end{figure}

\subsection{DQ Her superhump period}
     
\begin{table}
\caption{Prominent periods table for DQ Her periodogram of SuperWASP subset data in the region of 0.1 to 2 days with 2000 trial periods (refer to Fig 9).} 
\label{table:1}  
\centering       
\begin{tabular}{c c c c}     
\hline\hline
Peak  & Frequency    &  Period & Power \\    
 &  Cycles /day  & (d)   \\  \hline
 \hline   
  a &  5.16450 & 0.1936 & 171.1451  \\
  b &  6.16675 & 0.1622 & 136.9614  \\
  c &  4.16225 & 0.2403 & 134.5336  \\
  d &  9.33025 & 0.1072 & 116.6206  \\
  e &  7.16900 & 0.1039 & 68.9775   \\
  f &  8.32800 & 0.1201 & 65.3540   \\
  g &  3.17475 & 0.3160 & 64.2832   \\
\hline
\end{tabular}
\end{table}
Using the orbital period of 0.1936d, and calculating the apsidal superhump period excess $\varepsilon_+$ from the Montgomery (2001) relationship for apsidal precession, an estimate of a possible positive superhump period for DQ Her can be made.
In order to place some limits on the accuracy of the estimated positive superhump period and prograde disc precession period, the Montgomery relationship was calculated for the range of masses of DQ Her given by Horne et al (1993) as $M_1$ = 0.6 $\pm$ 0.07 M$_{\odot}$ and $M_2$ = 0.4 $\pm$ 0.05 M$_{\odot}$. 
This gave a minimum mass ratio q = 0.5244, a maximum q = 0.8496 leading to $\varepsilon_+$ lying between 0.185 and 0.318. The predicted positive superhump period would then lie between 0.229d and 0.288d with prograde disc precession periods between 0.616d and 1.240d.

The prominent periods table (Table 1) for the Lomb-Scargle analysis shown in Fig. 9 gives a peak close to that predicted above. However folding the data on the possible positive superhump period of 0.2403d, does not give much confidence that this value is in reality a superhump period.
Given an orbital period of 0.1936d and a taking $M_1$ = 0.6M$_{\odot}$ and $M_2$ = 0.4M$_{\odot}$ a positive superhump period of 0.2389d is predicted, and a likely prograde disc precession period of 1.013d. If this disc precession period is true, it is unlikely that the prograde disc precession period could be seen in these ground-based observations due to the sampling rate. No evidence was, therefore, found for a positive superhump period or prograde disc precession period in the data sets. 

    We also looked for negative superhumps by prewhitening the DQ Her periodogram of the SuperWASP data subset in the region of 0.1 and 2 days with the orbital period. This removed all periods other than the 0.1072d and the 0.1201d periods seen in Table 1. These, however, remained only at significantly reduced power levels. No obvious sign of negative superhumps could be seen at these periods. Retrograde disc precession periods were calculated  from the 0.1072d and 0.1201d potential negative superhump periods and searched for in the data set. None were found.

\subsection{Eclipse analysis of DQ Her SuperWASP data}

Clearly since no negative superhump period was found in the DQ Her data, there is no corresponding variation of eclipse depth with negative superhump phase as is seen in PX And. However we investigated whether there were any other modulations in the eclipse depths. The depth of eclipses in DQ Her were determined by the method described in Section 4.5. The reference flux determined for measuring the eclipse depth was itself found to be varying and showed a distinct trend in decreasing flux with increasing time. The reason for this is not immediately apparent.
This change in the reference flux clearly affects the determination of the eclipse depth and shows that as the  reference flux increases the measured depth of the eclipse also increases.  Accordingly the depth of the eclipse was measured as the ratio of the eclipse depth flux to the reference flux. 

The trend of the relative eclipse depth with orbital cycle number shown in Fig.10 suggests a potential periodic modulation of the eclipse depth around $\thicksim 600$ orbital cycles or, alternatively, 112 days. An illustrative sine curve with this period is shown overlaid on Fig.10 to aid the eye.
In order to see if this period was actually present in  a more extensive data set, a Lomb-Scargle analysis was performed in the region of 1 to 300 days using the complete SuperWASP data set of 4355 observations
The results of the Fourier analysis are shown in Fig.11. The most prominent period lies at 6.7183d, which is an observing alias, however there is a broad peak at roughly 100 days, which is close to the expected period of 112d from the eclipse depth analysis above. 
The spectral window for the 1 to 300 day period in the periodogram is flat with no significant prominent periods up to approximately 120 days at which point it gradually begins to slope upwards.
\begin{figure}
   \resizebox{\hsize}{!}{\includegraphics{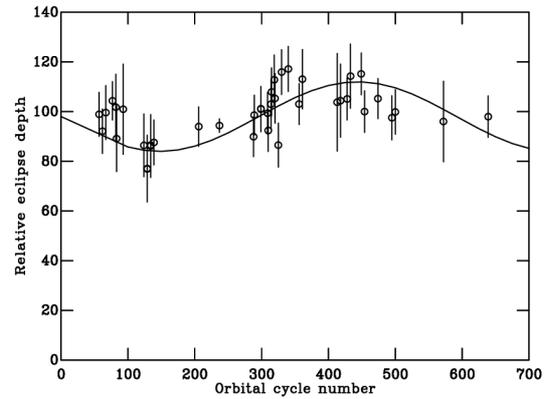}}
  \caption{DQ Her relative eclipse depth versus cycle number with an illustrative sine curve overlaid on the data to aid the eye.}
  \label{Fig.10}
\end{figure}
\begin{figure}
   \resizebox{\hsize}{!}{\includegraphics{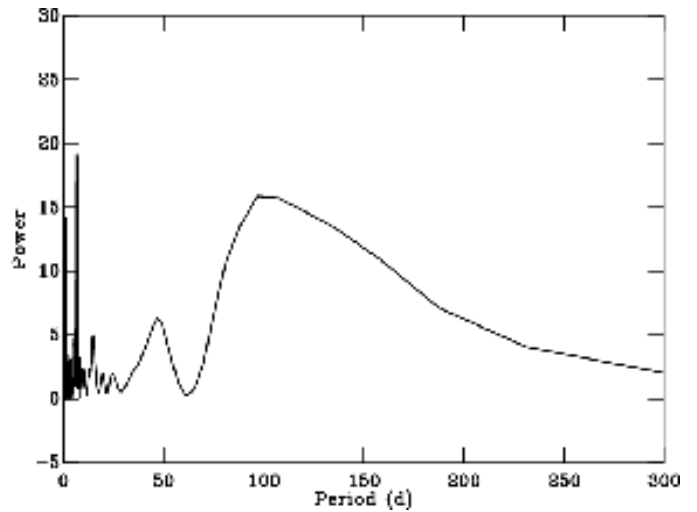}}
  \caption{Periodogram for DQ Her SuperWASP data spanning 1 to 300 days with 1500 trial periods.}
  \label{Fig.11}
\end{figure}
\subsection{ V795 Her}
In order to see what periods could be detected in the SuperWASP observations of V975 Her, a subset of the complete light curve data set was again chosen in order to minimise the number of sampling aliases, and a Lomb-Scargle analysis was performed. The analysis was carried out between 0.1d and 2.0d periods on the 2200 observations. The resulting periodogram shows that the dominant periods occur below 0.3d, but there was no sign of the expected prograde disc precession period of 1.53d. We investigated the 0.1d to 0.3d region at greater resolution. The resulting periodogram shows the dominant period is 0.1165d; this is the positive superhump period. There is no clear sign of the orbital period at 0.10826d.(\cite{Woodetal2009})  
Prewhitening a 0.1d to 0.2d periodogram with the positive superhump period removes all peaks above a power level of 5\% of the power level of the positive superhump period. i.e. the orbital period still cannot be seen and there are no other significant periods. 

\subsection{V975 Her Dynamic period analysis}
The superhumps in V795 Her were found to vary in amplitude and shape over the course of the observations. Figure 12 shows the evolution of the superhumps over the 150 day observational period.
In order to characterise this, we performed a dynamic period analysis on the data set. The SuperWASP data was separated into approximately 10-day observation sets.  The start of each observation set begins near the mid point of the previous set of 10 observations (ie at the overlap of 50\% of the previous observations) and a Lomb-Scargle analysis performed on each data set to obtain the power function of the 0.1164d positive superhump period. 
The resulting variation in the power function at the positive superhump period is shown in Fig.13.

\begin{figure}
   \resizebox{\hsize}{!}{\includegraphics{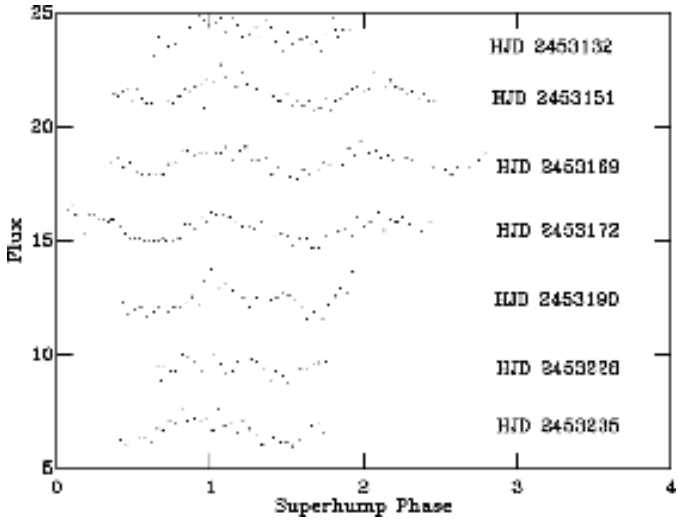}}
  \caption{A representative set of V795 Her SuperWASP light curves over the 150 day observation period  showing the evolution of the superhumps. Individual lightcurves are offset for clarity.}
  \label{Fig.12}
\end{figure}
\begin{figure}
   \resizebox{\hsize}{!}{\includegraphics{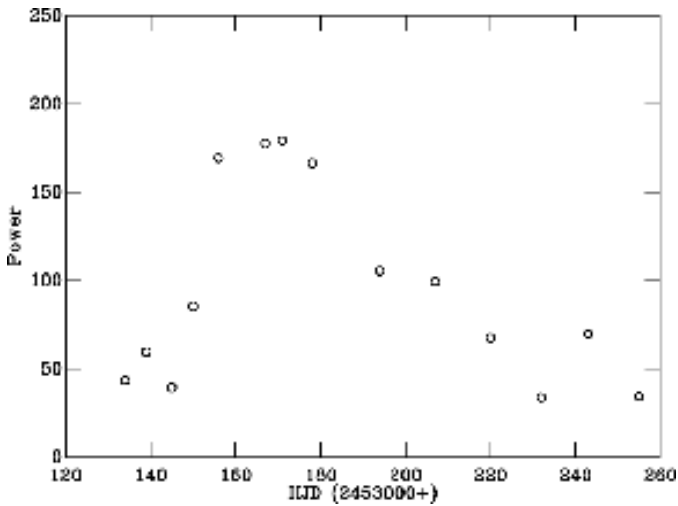}}
  \caption{Variation with time of the power in the superhump period for V795 Her.}
  \label{Fig.13}
\end{figure}

\section{Discussion}
\subsection{PX And discussion}
The SuperWASP observations show that the average white light magnitude for PX Andromedae is 14.88 mag.  When Green et al (\cite{Greenetal1982}) first observed it they quoted its magnitude as B=15.5. Thorstensen et al (\cite{Thorstensenetal1991}) found it’s magnitude to be V=16.9 $\pm 0.2$. It appears that the disc in PX And varies between high and low states, but there is no data indicating over what period this occurs.
As noted earlier, Patterson (1999) reported PX And to show positive superhumps with a superhump excess of $\varepsilon_+$ $\simeq 0.09$. Taking an orbital period of 0.1463d (Stanishev et al 2002) the relationship infers that a  positive superhump period in the region of 0.15946d should be seen. No significant periods in this region were detected in the period analysis of the SuperWASP data. It is possible therefore that the positive superhump period present in 1999 had disappeared in 2004 as PX And may have been in a low state. Patterson's (1999) observations also revealed PX And to show negative superhumps with a period of 0.1415d and a precession period of 4 to 5 days. Stanishev et al (\cite{Stanishev}), based on five observing runs during October 2000, identified the negative superhump period as 0.142d but were unable to make a reliable error estimate of the retrograde disc precession period because the data set was short and only covered approximately 1 cycle.
The results of the period analysis of the complete WASP data set of 126 days of observations in 2004 show that the negative superhump period is 0.1417 $\pm 0.0001$d and the retrograde disc precession period is 4.43 $\pm 0.05$d.
 
Stanishev et al (\cite{Stanishev}) also found a 0.207d periodicity, but they could not determine whether it was a real or a spurious signal. This period is seen here in Fig.2, but prewhitening the Fourier analysis with the superhump period, as shown in Fig.3, indicates that this period is likely to be an alias of the superhump period.

As reported by Thorstensen et al (1991), there is a considerable degree of variation in the shape and depth of the eclipses. Thorstensen et al (1991), using the 1.3m McGraw-Hill Telescope found the average depth of eclipse to be 0.5 mag. Stanishev (\cite{Stanishev}), using the 2.0m telescope at Rozhen Observatory detected an average eclipse depth of 0.89 mag. The results from the SuperWASP are found to be in good agreement with an average eclipse depth of 0.71 mag. 
Despite scatter in the SuperWASP data, the results of the period and eclipse analyses confirm Stanishev’s tentative conclusion that the eclipse depth in PX And is modulated with the 4.43d period corresponding to the retrograde precession of its accretion disc. The variation in relative eclipse depth with the superhump period found in this work is of the order of $\pm 0.05$ mag.

As noted in section 4.2, Stanishev et al reviewed the then current theories for the origin of the negative superhump light source but were unable to come to a satisfactory conclusion about how the eclipse depth varied with the negative superhump period. As part of their investigation they simulated eclipses of a tilted precessing disc. On the basis of the mean luminosity of a precessing disc which could give rise to the observed amplitude of the precessional modulation, Stanishev et al (\cite{Stanishev}) calculated the tilt of the disc in PX And to be between $2.5^{\circ}$ and $3^{\circ}$. Their  results showed that the eclipse depth did vary but only with a very low amplitude of 2-3\%, which was insufficient to explain the observations. 

It would appear that the simplest explanation for the variation of the eclipse depth in PX And is that the central portion of the disc is not eclipsed by the secondary star, but acts as an additional variable source of light which is modulated on the negative superhump period. As the negative superhump period of 0.1417d is shorter than the orbital period of 0.14635d, over the course of successive superhump cycles, the maxima in the negative superhump period progressively moves its position, compared with the timing of the eclipse in the orbital phase. Thus when the maxima of the negative superhump occurs simultaneously with the eclipse, the depth of the eclipse is reduced. This variation, repeats over $\sim31$ cycles of the negative superhump phase and is therefore modulated by the retrograde disc precession period of 4.43d. 

As noted in Section 1, the mechanism of the superhump light source itself, remains controversial. However, Montgomery (\cite{Montgomery2009a}), shows that a minimum disc tilt angle of around 4\degr is required to produce negative superhumps, close to that calculated for PX And. In addition, Montgomery (\cite{Montgomery2009a}), suggests that there may be two sources of the negative superhumps. One is gas stream overflow and the second is particle migration from the outer annuli into inner annuli of the disc. Both contribute to the negative superhump light signal. Therefore, when the gas stream is turned off as shown by Wood et al
(\cite{Woodetal2009}), only one source is eliminated, but not the second. The negative superhump will still persist in a tilted disc regardless if the stream is present or not and the source is from the centre of the disc as shown by Wood et al and Montgomery. These results would seem to support our explanation for the eclipse depth modulation in PX And.

\subsection{DQ Her discussion}

Fourier analysis of the SuperWASP DQ Her observations, 4400 data points taken over a 35 day period, unambiguously detected the orbital period of 0.1936d.
Calculation of the positive superhump period excess from the relationship derived by Montgomery (2001), using estimates of the white dwarf and secondary star masses from Horne et al (1993) resulted in predicted periods for possible apsidal superhumps lying between 0.229d and 0.288d with a possible prograde disc precession period between 0.616d and 1.24d. Fourier analysis of the data using the Lomb-Scargle method failed to find evidence of either positive superhumps or a prograde disc precession.

The most likely reason for the lack of evidence of apsidal superhumps in DQ Her may lie in the mass ratio of the system. Based on masses of the primary and secondary star in DQ Her taken from Horne et al (1993) the mass ratio is thought to be between 0.524 and 0.850. 

 Montgomery (2001) indicates that his relationship for the apsidal superhump period excess is only strictly valid for mass ratios $q < 0.33$. Theoretically positive superhumps are not believed to  develop in systems with high mass ratios (and by implication long orbital periods) as resonance is unable to set  in, and  as a consequence, the disc is not able to grow  enough to become eccentric. Based on a mass ratio of q = 0.5244, the tidal radius of the disc in DQ Her is expected to be 0.3935\emph{a} where \emph{a} is the binary separation distance. Using the formulae given in Hellier (\cite{Hellierbook}), the radius of the 3:1 inner Linblad resonance radii would be situated at 0.4177\emph{a} and thus it is unlikely that a stable resonance orbit would be maintained at that distance. However for 4:1 resonance, the stable orbit would be 0.3448\emph{a}, well within the tidal radius. DQ Her is an Intermediate Polar and it is likely that the inner disc will be truncated beyond the recircularisation radius of 0.113\emph{a} but how far into the disc, and whether the disruption reaches the 4:1 resonance distance remains an open question. It may be, that the magnetic field in DQ Her disrupts the accretion disc to such an extent that disc resonances do not establish themselves sufficiently and hence superhumps are not seen. 

Analysis of the eclipses in DQ Her suggests that there may be a possible long-term periodicity in the region of 112 days. Although the periodogram analysis does reveal a broad peak around a 100-day period, this cannot be reliably confirmed since the detected period is greater than half the period range of observations. The spectral window for this analysis is, unfortunately, also indeterminate.
This, however, leaves open the intriguing possibility that the 100 day periodicity is in fact real.  Richman et al (\cite{Richmanetal1994}), in examining long term periods in CVs, present an AAVSO light curve of the nova-like variable TT Arietis, for the period 1975 to 1992, in which a modulation over a 300 day period is clearly seen to be present. There is evidence that such long term oscillations may be associated with star spot activity. Warner \cite{Warner1988} has demonstrated that the variation of the quadropole moment, caused by solar-type magnetic cycles, produces cyclic variation of some observable parameters of CVs. Increasing the number of flux tubes causes an increase in the radius of the star (or a decrease in the  Roche lobe radius  as described by Richman et al (\cite{Richmanetal1994})), resulting in an enhancement of  mass transfer.  Enhanced mass transfer gives rise to an increase in the bright spot luminosity and mass transfer through accretion disc. Thus cyclical magnetic activity of the secondary star in a CV can be observed as cyclical variations in the brightness of the systems (Ak et al 2001). Longer term quasi-periodic variations are already known to exist in UX UMa on timescales between 7 and 30 years and similarly 6 to 14 years for DQ Her 
(Warner \cite{Warner1988}), which are believed to be associated with magnetic activity. Magnetic moments of secondary stars are thought to be able to affect mass transfer in the vicinity of the L1 point (eg Barrett et al \cite{Barrett}) and lead to observable changes in the orbital period and mean brightness of the system. For example in the case of DQ Her, such oscillations give rise to an O-C amplitude of 1.2 minutes (Rubenstein \cite{Rubenstein1991}) and there are 0.3 mag variations in brightness of the nova (Warner \cite{Warner1988}) over the 7 to 30 year quasi-period.  It is entirely feasible that the potential 100 day period indicated in this analysis, is associated with DQ Her's magnetic activity. We fully support the suggestion by Dai \& Qian (\cite{DaiQian}) that further detailed analysis of DQ Her's variation in brightness may help to clarify the role of magnetic activity in the observed modulations in DQ Her's O-C diagram.                  

\subsection{V795 Her discussion}
The periodogram analysis for V795 Her was able to detect the likely positive superhump period 0.1165d. However both the 0.10826d orbital period and the 1.53d prograde disc precession period were not seen.
Photometric observations by Shafter et al (\cite{Shafteretal1987}), Ashoka et al (1989), Klazuny et al (1989), Zhang et al (1991), Zwitter et al (\cite{Zwitteretal1994}) and this work based on 2004 SuperWASP data all detected periodic variations in the region of 0.1165d which is attributed to the presence of a positive superhump signal. No sign of the 0.10826d orbital period was seen in any of the above work. Spectroscopic observations on the other hand by Shafter et al (1990), Prinja et al (\cite{PrinjaRosen1993}), Zwitter et al  (\cite{Zwitteretal1994}), and Casares et al \cite{Casares1996} identified the 0.1082d period which is attributed to the orbital period.  However, Shafter et al (1990) were able to identify the orbital period in their photometric data, but the superhump period was no longer detected.
It would appear that the orbital period may only be seen in photometric data when the positive superhump activity is minimal.

The approximate 120 day modulation in the strength of the superhump period is very intriguing. What we are seeing in the dynamic power spectra is that there is a measurable variation in the coherence of the light output from the system at the period of superhumps, which is the “beat” period of the orbital motion of the secondary with the disc precession period. Comparison of the individual light curves against the corresponding Heliocentric Julian Date in the power spectrum in Fig.13 shows that while there is some variation in the form of the light curves the steep increase and then decline in the power spectrum does not appear to be reflected in corresponding changes in flux. This may be simply because each point in the power spectrum is an average of 10 days of observations, while Fig.12 shows single nightly light curves. Comparison of the flux amplitude in the light curves with the periodogram power function also is inconclusive.

It is difficult to speculate, at this point, on the cause of the modulation in the power spectra, since the length of the period is similar to the length of the period of observations. Therefore we cannot be sure whether the power spectra period is stable or whether it varies with time.As further long term SuperWASP data on V795 Her become available, it is proposed to follow up these observations.

\subsection{Conclusions}

We have undertaken periodogram analyses on long term SuperWASP observations of PX And, DQ  Her, and V795 Her. We confirm the negative superhump period for PX And is 0.417 $\pm$ 0.0001d. We find no evidence for positive superhumps during these observations. Despite the low signal to noise in the data giving rise to variations in the determination of eclipse depths in both PX And and DQ Her, we found  the 4.43 $\pm$ 0.05d retrograde disc precession in PX And to modulate the depth of its eclipses, as suggested by Stanishev et al (\cite{Stanishev}). Our observations of DQ Her found no evidence of either positive or negative superhumps or either a prograde or retrograde disc precession. It would appear that eclipse phase modulations are not related to positive superhumps and their associated  prograde precession periods. If eclipse phase modulations are associated with disc tilt, then positive superhumps and their prograde precession periods do not appear to be associated with disc tilt. In DQ Her we find evidence for a modulation of the eclipse depth over a period of 100 days which may be linked with solar-type magnetic cycles which give rise to long term photometric variations. The periodogram analysis for V795 Her detected the likely positive superhump period 0.1165d, however, neither the 0.10826d orbital period or the prograde 1.53d disc precession period were seen. We did however found a variation in the periodogram power function at the positive superhump period, over a period of at least 120 days. The cause of this variation is presently unknown and is to be further investigated.

\begin{acknowledgements}
     This research has made use of NASA's Astrophysics Data System and the code used in the peridogram analysis was supplied through the Centre for Backyard Astrophysics Belgium observatory. The WASP Consortium comprises astronomers primarily from the Universities of Keele, Leicester, The Open University, Queen’s University Belfast and St Andrews, the Isaac Newton Group (La Palma), the Instituto de Astrof´ısica de Canarias (Tenerife) and the South  African Astronomical Observatory. The SuperWASP-N camera is hosted by the Isaac Newton Group on La Palma with funding from the UK Science and Technology Facilities Council. We extend our thanks to the Director and staff of the Isaac Newton Group for their support of SuperWASP-N operations. We would also like to thank the anonymous referee for helpful suggestions and useful comments made in the preparation of this paper.

\end{acknowledgements}

\end{document}